\documentclass[12pt,preprint]{aastex}
\shorttitle{Accurate VLBI Astrometry of PSR J0437--4715}
\shortauthors{Deller, Verbiest, Bailes \& Tingay}

\begin{document}

\newcommand{\pbdot}{\ensuremath{\dot{P}_{\mathrm b}}}
\newcommand{\Gdot}{\ensuremath{\dot{G}}}
\newcommand{\degrees}{\ensuremath{^{\circ}}}

\title{Extremely high precision VLBI astrometry of PSR J0437--4715 and implications for 
theories of gravity}

\author{A.T. Deller\altaffilmark{1,2}, J.P.W. Verbiest\altaffilmark{1,2},  S.J. Tingay\altaffilmark{3} \& M. Bailes\altaffilmark{1}}
\altaffiltext{1}{Centre for Astrophysics and Supercomputing, Swinburne University of Technology, Mail H39, P.O. Box 218, Hawthorn, VIC 3122, Australia}

\altaffiltext{2}{co-supervised through the Australia Telescope National Facility, P.O. Box 76, Epping, NSW 1710, Australia}

\altaffiltext{3}{Department of Imaging and Applied Physics, Curtin University of Technology, Bentley, WA, Australia}

\begin{abstract}
Using the recently upgraded Long Baseline Array, we have measured the trigonometric parallax
of PSR J0437--4715 to better than 1\% precision, the most precise pulsar
distance determination made to date.  Comparing this VLBI distance measurement to
the kinematic distance obtained from pulsar timing, which is calculated from the pulsar's 
proper motion and apparent rate of change of orbital period,
gives a precise limit on the unmodeled relative acceleration between the Solar System
and PSR J0437--4715, which can be used in a variety of applications. Firstly,
it shows that Newton's gravitational constant $G$ is stable with time
($\Gdot/G  = (-5 \pm 26) \times 10^{-13}$\ yr$^{-1}$, 95\% confidence). Secondly,
if a stochastic gravitational wave background existed at the 
currently quoted limit, this null result would fail
$\sim\!50$\% of the time. Thirdly, it excludes Jupiter-mass planets within 226 AU
of the Sun in 50\% of the sky (95\% confidence). Finally, the $\sim\!1$\% agreement of the parallax
and orbital period derivative distances provides a fundamental confirmation
of the parallax distance method upon which all astronomical distances are
based.
\end{abstract}


\section{Introduction}
PSR J0437--4715 has been observed intensively since its discovery by \citet{johnston93a}.
It is the brightest and nearest observed millisecond pulsar, and has also been 
studied in the optical \citep{bell93a}, ultraviolet \citep{kargaltsev04a}, and X--ray 
\citep{zavlin02a} bands.  The high rotational stability and close
proximity of this pulsar--white dwarf binary system make it an excellent
probe of General Relativity (GR) and alternative forms of gravitational theories.  The measurement
of its Shapiro delay by \citet{van-straten01a}, where the radio waves from the pulsar are delayed
as they pass through the gravitational potential of the companion, is one such test which has
shown consistency with GR predictions.  The search for the low frequency stochastic gravitational
wave background (GWB) using pulsar timing arrays 
\citep[e.g.][]{jenet05a} is another test of GR which is 
facilitated in part by timing of J0437--4715.

Although variation of Newton's gravitational constant $G$\ with time is forbidden in GR, 
this is not required in alternate formalisms of gravity \cite[e.g.][]{brans61a}.
\citet{verbiest08a} have measured the rate of change of orbital period \pbdot\ in the 
J0437--4715 system
to better than 2\% precision and shown that the agreement of the derived kinematic distance with  
the parallax distance derived from timing limits the time variation of  
$G$\ to less than 3 parts in $10^{11}$.
 In this Letter, we present a new Very Long Baseline Interferometry (VLBI) determination
of the position, proper motion, and parallax of PSR J0437--4715 and show that this improves the
previously published \Gdot\ limit derived from this system by an order of magnitude, approaching the
best published limit ($\Gdot/G  = (4\pm9)\times 10^{-13}$\ yr$^{-1}$), which is derived from 
Lunar Laser Ranging 
\citep[LLR;][]{williams04a}.  Our VLBI observations and results are presented in 
\S\ref{sec:vlbi} and \S\ref{sec:results} respectively and the limitations on apparent accelerations
(due to \Gdot\ or other possible causes such as unseen massive planets) are 
derived in \S\ref{sec:anomacc}.
In \S\ref{sec:gwb}, we investigate the impact of the stochastic GWB
on our results and derive an independent
limit on the GWB amplitude.  We summarize our conclusions in \S\ref{sec:conclusions}.

\section{VLBI observations}
\label{sec:vlbi}
Details of the observational, correlation, and post--processing strategies used are
covered in detail in \citet{deller08a} and only a brief summary is presented here.
Four observational epochs (MJD 53868, 54055, 54181, 54416) 
spanned 1.5 years, with each epoch lasting 12 hours and utilizing
all available antennas from the Australian Long Baseline Array (LBA).  
These were the three ATNF antennas (ATCA, Parkes,
Mopra), the two University of Tasmania antennas (Hobart and Ceduna) and, when available,
one of the NASA Deep Space Network (DSN) antennas at the Tidbinbilla 
station\footnote{DSS43 (70m) participated on MJD 54055 and 54181, while
DSS34 (34m) was used on MJD 54416}.  Observations were made at 8.4 GHz, and
four 16 MHz bands were used, quantized at two bit precision to give a total data rate of 256 Mbps.

Phase referencing was conducted using the International Celestial Reference Frame
(ICRF) source J0439--4522, a bright (several hundred mJy flux at 8 GHz) and compact radio 
source less than 2\degrees\ from PSR J0437--4715, whose position is known to 
1 mas \citep{ma98a}.  A six minute target/calibrator cycle 
was employed, with observing time divided equally between the target 
and calibrator.  The well--studied blazar PKS 0537--441 was used for bandpass calibration.

The data were correlated, using matched filtering (gating) on pulse profiles, with the DiFX 
software correlator \citep{deller07a}.  Due to the extremely narrow pulse profile of PSR J0437--4715
at 8.4 GHz, this gating improves the output signal to noise ratio (SNR) by 
a factor of 7 compared to ungated data.  The pulsar ephemeris used to set gates was taken from
\citet{hotan06a}.

Due to the relatively high frequency used for these observations, several of the calibration steps 
described in \citet{deller08a} proved unnecessary.  Data reduction was performed with and
without GPS--based ionospheric correction, and the small ($\sim100\,\mu$as) corrections which
resulted from ionospheric correction gave a poorer parallax fit overall, with double the
estimated systematic errors.  Given the uncertainty of 
GPS--based corrections at these Southern latitudes, the ionospheric corrections were therefore
not applied.  Additionally, the effect of scintillation was
negligible at these higher frequencies, so the scintillation correction detailed in 
\citet{deller08a} was not applied.

Attaining single epoch accuracies of $\sim\!100\,\mu$as, as shown in \S\ref{sec:results},
required post-correlation corrections to the visibilities due to the proper motion of the pulsar
($\sim\!200\,\mu$as in each 12 hour observation).  Additionally, the orbital motion of
PSR J0437--4715 causes a peak to peak displacement of $\sim\!110\,\mu$as.
Compensation for this displacement in the single--epoch positions
improved the quality of the proper motion and parallax fit, reducing the estimated systematic
error contribution by 3\%.

\section{Astrometric results}
\label{sec:results}
As discussed in \citet{deller08a}, the optimal weighting for visibility points when imaging 
the pulsar to fit for position depends on the ratio 
of thermal noise to residual systematic errors.  For strong targets systematic errors dominate the
error budget, meaning the use of equally weighted visibilities (as opposed to weighting visibilities by
baseline sensitivity) produces superior results.  For PSR J0437--4715 the SNR of a typical
single--band, single--epoch detection was $\sim\!20$, well above the threshold of 
$\sim\!10$\ identified in \citet{deller08a}, and thus we would expect equally weighted
visibilities to produce superior results.  To confirm this, the parallax was determined
from two separate datasets, which were produced using sensitivity--weighted 
and equally weighted visibilities.  As expected,
the equally weighted dataset produced a better fit, with smaller errors on fitted parameters and
smaller estimated systematic error contributions.

\begin{figure}
\begin{center}
\begin{tabular}{cc}
\includegraphics[width=0.4\textwidth]{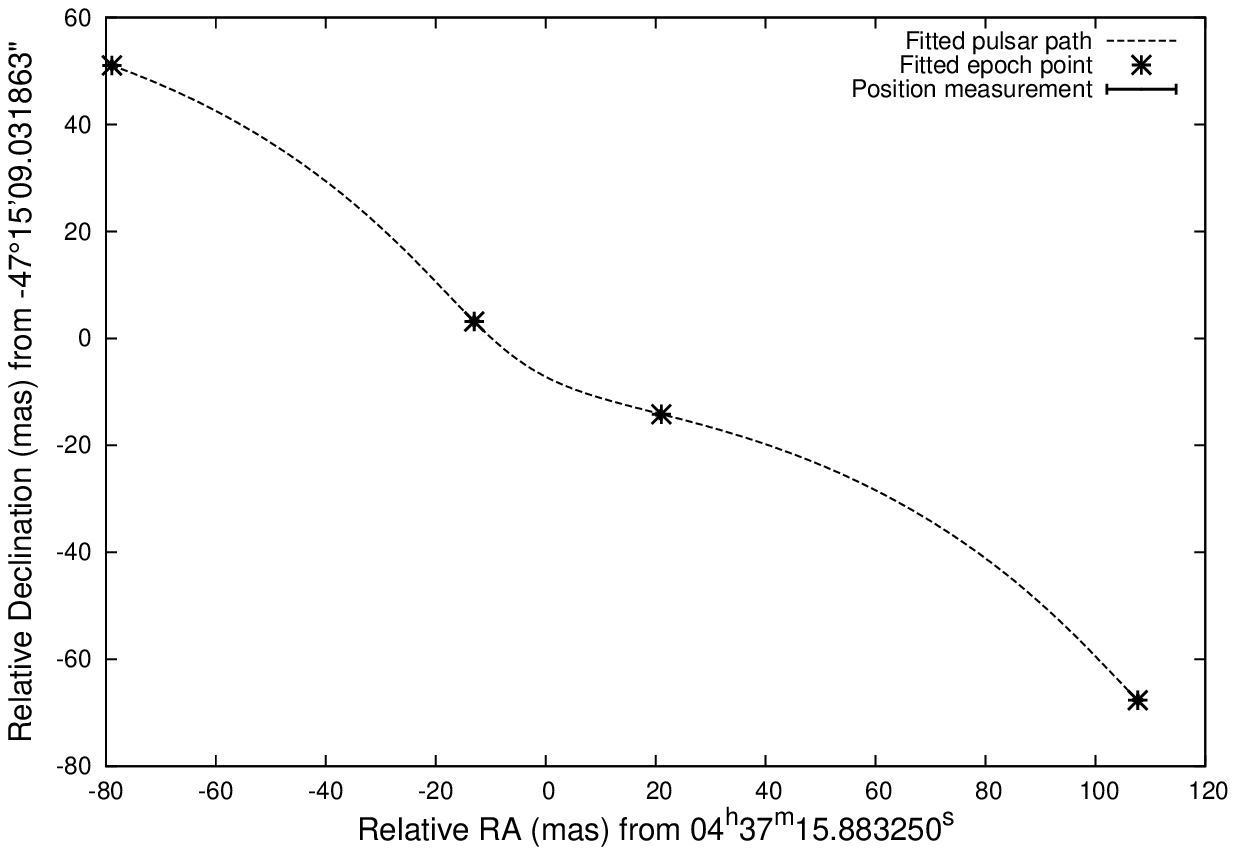} &
\includegraphics[width=0.4\textwidth]{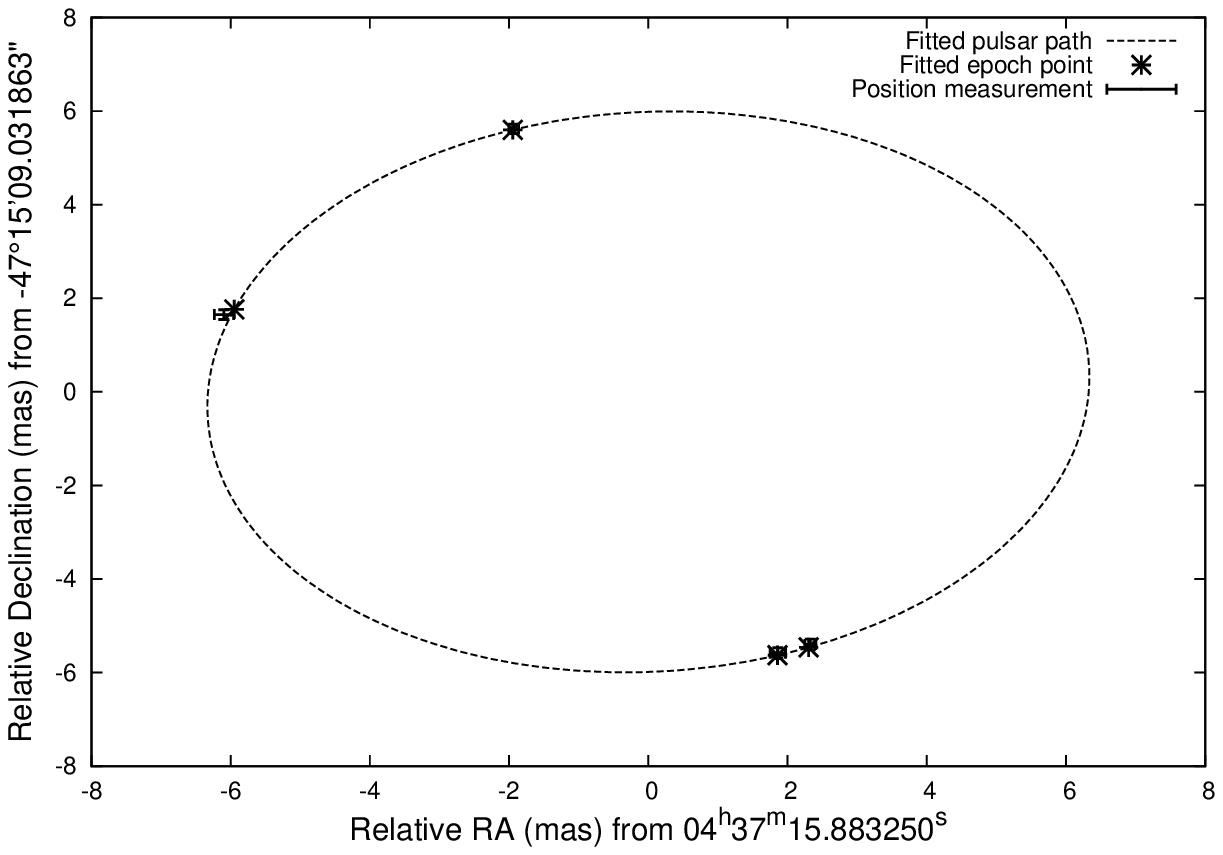} \\
\includegraphics[width=0.4\textwidth]{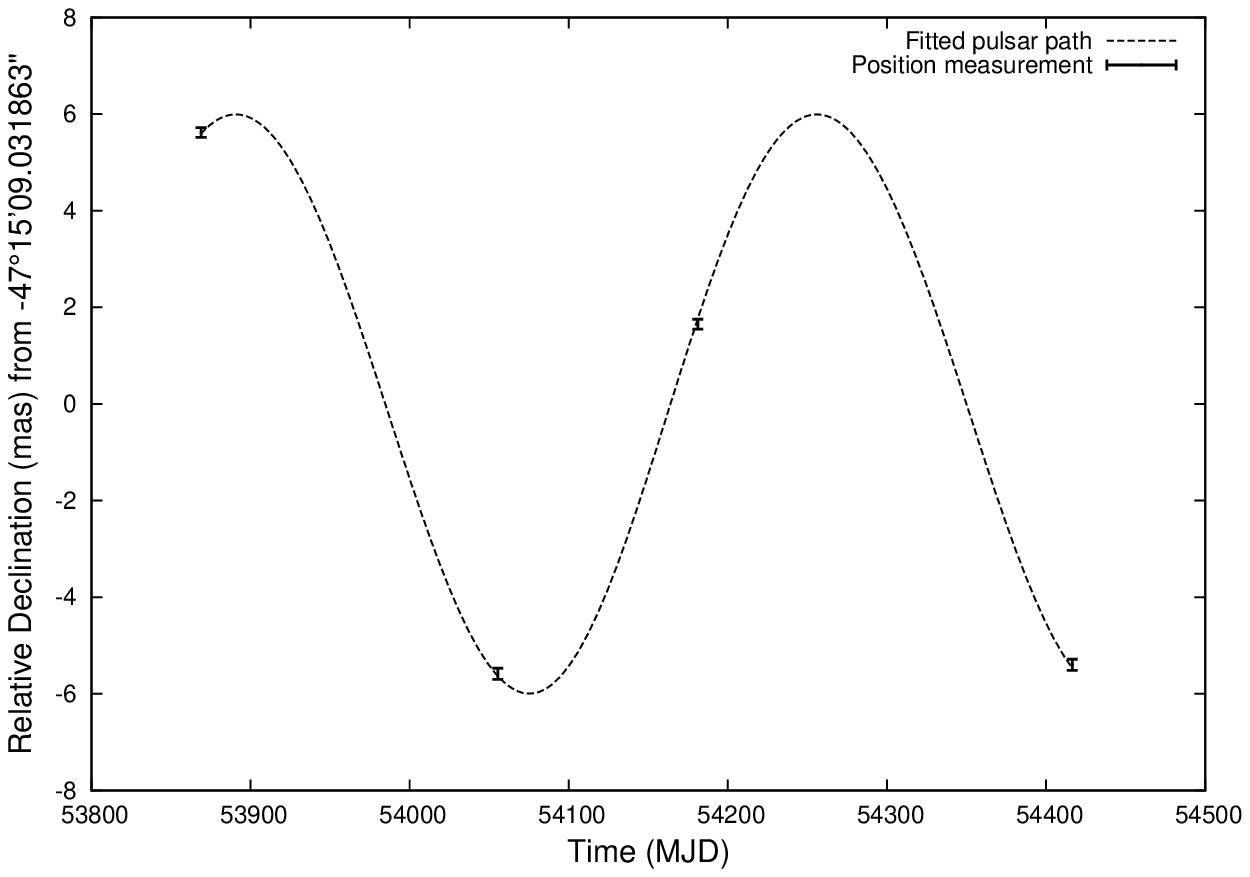} &
\includegraphics[width=0.4\textwidth]{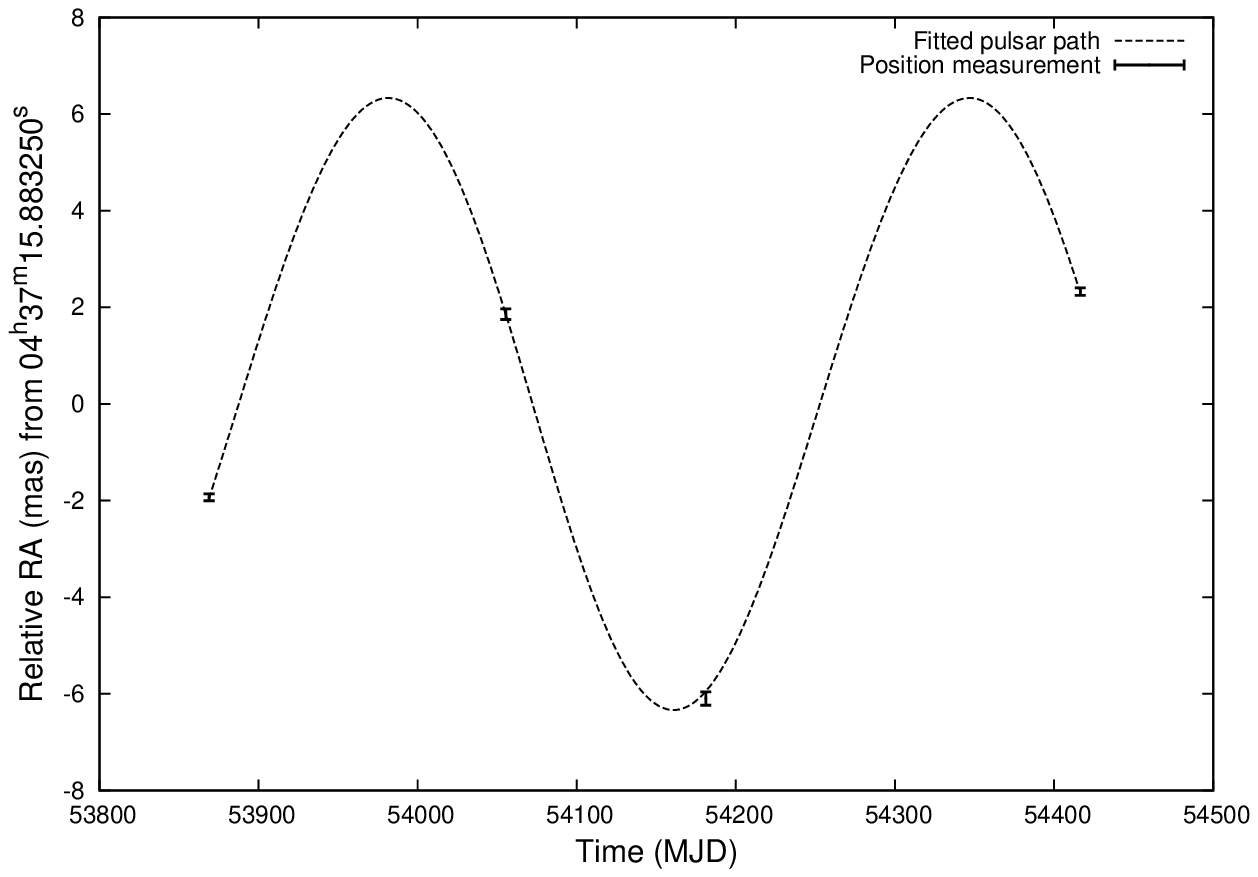} \\
\end{tabular}
\caption{Motion of PSR J0437--4715, with measured positions
overlaid on the best fit.  Clockwise from top left: Motion in declination vs right ascension; as before 
but with proper motion subtracted; right ascension vs time with proper motion subtracted; and
declination vs time with proper motion subtracted.}
\label{fig:0437radec}
\end{center}
\end{figure}

The data were fitted using an iterative approach
designed to estimate and account for systematic errors, described in \citet{deller08a}.  The
resulting parameter values are shown in Table~\ref{tab:results} along with the \citet{verbiest08a}
timing measurements\footnote{The timing positions of \citet{verbiest08a} have been re--fitted
at the VLBI reference epoch of MJD 54100}; 
the VLBI fits to pulsar motion
and observed positions are shown in Figure~\ref{fig:0437radec}.  
All errors are 1$\sigma$\ unless otherwise stated, and include the covariances between fitted 
parameters.  Over this short timespan, the covariance between parallax and proper motion is 
significant, and constitutes approximately half of the quoted parallax uncertainty.

The average total single--epoch error of 132\,$\mu$as is the best relative astrometry performed 
by the LBA to date, and is similar to other recent VLBI astrometric results \citep[e.g][]{loinard07a}.  
By way of comparison, the limiting accuracy for EVN observations at 8.4 GHz has been simulated 
by \citet{pradel06a} to be 83\,$\mu$as with a 1\degrees\,calibrator--target separation, at a 
declination of 50\degrees.  The EVN has similar sensitivity to the LBA, but
somewhat better uv coverage and longer baselines.

Our derived distance of $156.3 \pm 1.3$\,pc is the most accurate distance
measurement (in both absolute and fractional distance) 
for a pulsar to date and approaches the most accurate distance measurements 
made of objects outside the solar system \citep[T Tauri, $147.6 \pm 0.6\,{\mathrm{pc}}$:][]{loinard07a}.  
Previously, the highest--precision VLBI pulsar distance determinations were 
those made by \citet{brisken02a}, who measured the distance of PSR J0953+0755 to 
an accuracy of 5 pc, along with 8 other Northern Hemisphere pulsars.
The two previous parallax measurements made using a Southern array, of PSR J0835--4510 
\citep{dodson03a} and PSR J1456--6843 \citep{bailes90a}, had 1$\sigma$\,distance
errors of 19 and 70 pc respectively.  

\subsection{Comparison to timing astrometry}

To compare the VLBI and timing positions, we have re--fitted the timing data
of \citet{verbiest08a} to obtain the position of 
PSR J0437--4715 at our reference epoch (MJD 54100).
Table~\ref{tab:results} shows that the timing and VLBI positions at MJD 54100 differ by
over two mas, many times the formal errors shown. However, the formal errors are negligible
compared to the 1 mas uncertainty in the VLBI phase reference calibrator position
and the potential constant offsets due to phase--referencing
errors such as station position errors \citep[known to exist at the cm level for the LBA;][]{deller08a}.  
Discrepancies between interferometric and timing positions of even larger magnitudes have
been found using the DE200 frame \citep{bartel96a}, 
and differences at the mas level still exist for the position
PSR J0437--4715 calculated using the newer DE414 solar system ephemeris, as compared to the
DE405 ephemeris used by \citet{verbiest08a}. Thus, we conclude that the VLBI and timing 
position difference is consistent with the uncertainty in the calibrator position 
and the offset between the solar system frame and the ICRF.

The values obtained for the VLBI proper motion differ by $\sim4\sigma$\ in both right
ascension and declination from the timing values.  A likely cause for this discrepancy is small 
changes in the centroid position of the phase reference source due to intrinsic source variability, 
which would be absorbed into the proper motion fit.  The phase reference source, which is
modeled by a $\sim1\,$mas FWHM Gaussian with additional delta components within
5 mas of strength 2\% and 0.2\% of peak flux, 
shows no gross evidence of variability (width/positions of primary/secondary components
constant to $\sim200\,\mu$as, and secondary fluxes are constant to $\sim1\,$mJy), but as it is only 
barely resolved (the beamsize is $\sim3\,$mas) variability at the $\sim100\,\mu$as level 
would be difficult to detect.  \citet{titov07a} show that some ICRF sources exhibit apparent
proper motions of hundreds of $\mu$as yr$^{-1}$ for periods of several years, 
and while a detailed VLBI history of this source is 
not available, calibrator measurements from ATCA show that the flux density has declined by a 
factor of three over a five year period, implying some variability.  
The higher proper motion precision obtained with the timing data 
(a factor of 5--7 times better than the VLBI results) reflects the fact that the timing data spans a 
time baseline 7 times longer than the VLBI dataset.

The parallax value obtained from VLBI is consistent with that derived from timing, and yields a
distance which is consistent with the kinematic distance of $157.0 \pm 2.4$\ pc 
derived from the orbital period derivative \pbdot.  However, the
VLBI parallax measurement is an order of magnitude more precise than the timing measurement,
and yields a distance which is a factor of two more precise than the kinematic distance.

\section{Limits on anomalous accelerations}
\label{sec:anomacc}
The newly measured parallax of $\pi = 6.396 \pm 0.054$\,mas allows an improved 
measurement of any anomalous acceleration of either the Solar System or 
PSR J0437--4715. Specifically, the apparent acceleration due to time variability of 
Newton's gravitational constant $G$ and the mass of an undetected trans-Neptunian 
planet near the line of sight to the pulsar can be limited.

\subsection{Constraints on \Gdot}
\label{subsec:gdot}
As first described by \citet{damour91a}, a precise measurement of a binary pulsar's orbital 
period derivative, \pbdot, can be used to constrain a variation of the 
gravitational constant, $G$. However, as \citet{bell96a} pointed out, for 
PSR J0437$-$4715 a precise distance needs to be known in order to 
correct \pbdot\ for the Shklovskii acceleration caused by its proper motion
\citep{shklovskii70a}. 
This analysis has been performed, based exclusively on timing data, by
\citet{verbiest08a}, whose limit was dominated by the uncertainty in their 
parallax measurement. Our VLBI parallax value, however, improves this limit by 
a factor of nearly 10 down to $\dot{G}/G = (-5 \pm 26)\times 10^{-13}$\ yr$^{-1}$\ at 
95\% certainty. This value compares well to the most stringent limit currently 
published: $(4\pm 9) \times 10^{-13}$\ yr$^{-1}$, derived through lunar laser 
ranging \citep{williams04a}. Since $\pi$ and \pbdot\ are now both measured to similar
precision, both measurements will have to be improved for a further 
significant increase in \Gdot\ sensitivity.  Additional VLBI observations and continued timing
could see this limit improve upon the existing LLR limit early in the next decade.

\subsection{Acceleration due to massive bodies}

An alternative source of anomalous acceleration is heavy planets in a 
wide orbit around the Sun or the pulsar. Building upon the initial analysis of 
\citet{zakamska05a}, our parallax measurement can be combined with the timing results 
from \citet{verbiest08a} to derive the following result: 
$a_{\odot}/c = (3 \pm 16 ) \times 10^{-20}\,{\mathrm s}^{-1}$\ at the $2 \sigma$ level. 
This improves the limit published in \citet{verbiest08a} by an order of magnitude and 
makes PSR J0437--4715 a more sensitive Solar System accelerometer than PSR J1713+0747, 
the most precise pulsar listed by \citet{zakamska05a}. From this, the limit for 50\% of the 
sky\footnote{within 60\degrees\,of the line of sight towards and away from PSR J0437--4715} 
can be calculated as: 
$|a_{\odot, 50\% {\mathrm{sky}}}/c| \leq 3.9\times 10^{-19}\,{\mathrm s}^{-1}$ (95\% certainty).
This acceleration limit can be used to exclude massive bodies within
a given radius of the Sun; for example, at Kuiper--belt radii (50 AU) it
excludes a planet more massive than Uranus over 50\% of the sky, while  
Jupiter--mass planets are excluded within 226\,AU over 50\% of the sky.

\section{Impact of the stochastic GWB}
\label{sec:gwb}

Using tools recently developed by \citet{hobbs08a}, we have simulated the effect of a GWB with spectral index $-2/3$ (as predicted for GWBs caused by black hole--black hole mergers) and dimensionless amplitude $1.1\times10^{-14}$\ (the best published GWB limit; \citealt{jenet06a}) on the observed value of \pbdot\ from pulsar timing. The simulated GWB causes the kinematic distance to be 
inconsistent with the VLBI parallax distance at the 2$\sigma$\, level in 
$\sim50\%$ of trials.  Thus, although these observations cannot improve 
upon the present GWB limit, they are consistent with it.
Simulations with a GWB with amplitude of $1.1\times 10^{-13}$ show inconsistencies between the kinematic and VLBI distances at the $2\sigma$ level in $95\%$\ of trials, providing an
independent exclusion of a GWB with an amplitude at or above this value.

It is also interesting to note that the precise limit on $\dot{G}$\ presented in \S\ref{subsec:gdot} would be impossible in a Universe with a strong GWB. In the simulations with GWB amplitude $1.1\times 10^{-13}$, the observed \Gdot\ value is inconsistent with 0 in $99\%$\ of cases, merely due to the GWB-induced corruption of the timing measurements.  Thus, the stochastic GWB must eventually limit the accuracy of measurements of \Gdot\ in the fashion outlined in this paper.

\section{Conclusions}
\label{sec:conclusions}
We have observed PSR J0437--4715 using the LBA and obtained the most precise pulsar 
distance determination to date, with an error $<1.5$\,pc.  Combined with accurate timing data, this
has enabled us to confirm that  $|\Gdot/G| < 3.1 \times 10^{-12}$\ yr$^{-1}$, 
the most stringent limit not obtained though
Solar System tests. Alternatively, assuming an unchanging gravitational force, the 
results can be interpreted as excluding any unseen Jupiter--mass planets within 226 AU of the Sun
in 50\% of the sky.
Finally, the agreement between VLBI and timing results in this single case implies an upper limit to the 
stochastic GWB amplitude which is within an order of magnitude of 
the best limit derived from observations of ensembles of pulsars.

\acknowledgements

This work has been supported by the Australian Federal Government's Major National Research Facilities program.  ATD is supported via a Swinburne University of Technology Chancellor's Research Scholarship and a CSIRO postgraduate scholarship.  The Long Baseline Array is part of the Australia Telescope which is funded by the Commonwealth of Australia for operation as a National Facility managed by CSIRO.

\clearpage

\begin{deluxetable}{lrr}
\tabletypesize{\tiny}
\tablecaption{Fitted VLBI results for PSR J0437--4715 and comparative timing values (positions
re--referenced to the VLBI proper motion epoch)}
\tablewidth{0pt}
\tablehead{
\colhead{Parameter} & \colhead{Fitted value and error} & \colhead{\citet{verbiest08a} timing values\phm{\tablenotemark{A}}}}
\startdata
Right Ascension (J2000)	&  	04$^{\mathrm h}$37$^{\mathrm m}$15.883250$^{\mathrm s}$ $\pm$ 0.000003	& 
				04$^{\mathrm h}$37$^{\mathrm m}$15.883185$^{\mathrm s}$ $\pm$ 0.000006\phm{\tablenotemark{A}} 	\\
Declination (J2000) 	&  	$-$47\degrees15'09.031863" $\pm$ 0.000037	& 
				$-$47\degrees15'09.034033" $\pm$ 0.000070\phm{\tablenotemark{A}}	\\
$\mu_{\alpha}$	(mas/yr)	&  	121.679 $\pm$ 0.052\phn\phn\phn	&
						121.453 $\pm$ 0.010\phn\phn\phn\phm{\tablenotemark{A}}	\\
$\mu_{\delta}$	(mas/yr)	&  	$-$71.820 $\pm$ 0.086\phn\phn\phn	&
						$-$71.457 $\pm$ 0.012\phn\phn\phn\phm{\tablenotemark{A}}	\\
Parallax $\pi$ (mas)	 		&  	6.396 $\pm$ 0.054\phn\phn\phn	&
						6.65 $\pm$ 0.51\phn\phn\phn\phn\phm{\tablenotemark{A}}	\\
Distance (pc)	&  	156.3 $\pm$ 1.3\phn\phn\phn\phn\phn		&
				157.0 $\pm$ 2.4\tablenotemark{A}\phn\phn\phn\phn\phn\\		
Transverse velocity $v_{\mathrm T}$ (km/s)	&  	104.71 $\pm$ 0.95\phn\phn\phn\phn	&
				104.9 $\pm$ 1.6\tablenotemark{A}\phn\phn\phn\phn\phn	\\
Reduced chi--squared 	&  	1.0 \phs \phn\phd\phn\phn\phn\phn\phn\phn$\  $ &
					 \phs \phn\phd\phn\phn\phn\phn\phn\phn$\  $\phm{\tablenotemark{A}}\\
Average epoch mean fit error (mas)			& 0.059 \phs \phn\phd\phn\phn\phn\phn\phn\phn$\  $ &
					 \phs \phn\phd\phn\phn\phn\phn\phn\phn$\  $\phm{\tablenotemark{A}}\\
Average intra--epoch systematic error (mas) 	& 0.068 \phs \phn\phd\phn\phn\phn\phn\phn\phn$\  $ &
					 \phs \phn\phd\phn\phn\phn\phn\phn\phn$\  $\phm{\tablenotemark{A}}\\
Average inter--epoch systematic error (mas) 	& 0.103 \phs \phn\phd\phn\phn\phn\phn\phn\phn$\  $  &
					 \phs \phn\phd\phn\phn\phn\phn\phn\phn$\  $\phm{\tablenotemark{A}}\\
Reference epoch for proper motion	(MJD)	& 54100.0 \phs \phn\phd\phn\phn\phn\phn\phn\phn$\  $&
			54100.0 \phs \phn\phd\phn\phn\phn\phn\phn\phn$\  $\phm{\tablenotemark{A}}\\
\enddata
\tablenotetext{A}{Derived from the kinematic distance obtained from \pbdot, 
not the less precise parallax values}
\label{tab:results}
\end{deluxetable}

\clearpage

\clearpage

\begin{thebibliography}{25}
\expandafter\ifx\csname natexlab\endcsname\relax\def\natexlab#1{#1}\fi

\bibitem[{{Bailes} {et~al.}(1990){Bailes}, {Manchester}, {Kesteven}, {Norris},
  \& {Reynolds}}]{bailes90a}
{Bailes}, M., {Manchester}, R.~N., {Kesteven}, M.~J., {Norris}, R.~P., \&
  {Reynolds}, J.~E. 1990, \nat, 343, 240

\bibitem[{{Bartel} {et~al.}(1996){Bartel}, {Chandler}, {Ratner}, {Shapiro},
  {Pan}, \& {Cappallo}}]{bartel96a}
{Bartel}, N., {Chandler}, J.~F., {Ratner}, M.~I., {Shapiro}, I.~L., {Pan}, R.,
  \& {Cappallo}, R.~J. 1996, \aj, 112, 1690

\bibitem[{{Bell} \& {Bailes}(1996)}]{bell96a}
{Bell}, J.~F., \& {Bailes}, M. 1996, \apjl, 456, L33+

\bibitem[{{Bell} {et~al.}(1993){Bell}, {Bailes}, \& {Bessell}}]{bell93a}
{Bell}, J.~F., {Bailes}, M., \& {Bessell}, M.~S. 1993, \nat, 364, 603

\bibitem[{Brans \& Dicke(1961)}]{brans61a}
Brans, C., \& Dicke, R.~H. 1961, Phys. Rev., 124, 925

\bibitem[{{Brisken} {et~al.}(2002){Brisken}, {Benson}, {Goss}, \&
  {Thorsett}}]{brisken02a}
{Brisken}, W.~F., {Benson}, J.~M., {Goss}, W.~M., \& {Thorsett}, S.~E. 2002,
  \apj, 571, 906

\bibitem[{{Damour} \& {Taylor}(1991)}]{damour91a}
{Damour}, T., \& {Taylor}, J.~H. 1991, \apj, 366, 501

\bibitem[{{Deller} {et~al.}(2007){Deller}, {Tingay}, {Bailes}, \&
  {West}}]{deller07a}
{Deller}, A.~T., {Tingay}, S.~J., {Bailes}, M., \& {West}, C. 2007, \pasp, 119,
  318

\bibitem[{{Deller} {et~al.}(2008){Deller}, {Tingay}, \& {Brisken}}]{deller08a}
{Deller}, A.~T., {Tingay}, S.~J., \& {Brisken}, W.~F. 2008, \apj, submitted

\bibitem[{{Dodson} {et~al.}(2003){Dodson}, {Legge}, {Reynolds}, \&
  {McCulloch}}]{dodson03a}
{Dodson}, R., {Legge}, D., {Reynolds}, J.~E., \& {McCulloch}, P.~M. 2003, \apj,
  596, 1137

\bibitem[{{Hobbs} {et~al.}(2008){Hobbs}, {Jenet}, {Lee}, {Verbiest}, {Yardley},
  {Manchester}, {Lommen}, {Coles}, {Edwards}, \& {Shettigara}}]{hobbs08a}
{Hobbs}, G.~B., {Jenet}, F.~A., {Lee}, K.~J., {Verbiest}, J.~P.~W., {Yardley},
  D., {Manchester}, R.~N., {Lommen}, A., {Coles}, W.~A., {Edwards}, R.~T., \&
  {Shettigara}, C. 2008, MNRAS, submitted

\bibitem[{{Hotan} {et~al.}(2006){Hotan}, {Bailes}, \& {Ord}}]{hotan06a}
{Hotan}, A.~W., {Bailes}, M., \& {Ord}, S.~M. 2006, \mnras, 369, 1502

\bibitem[{{Jenet} {et~al.}(2005){Jenet}, {Hobbs}, {Lee}, \&
  {Manchester}}]{jenet05a}
{Jenet}, F.~A., {Hobbs}, G.~B., {Lee}, K.~J., \& {Manchester}, R.~N. 2005,
  \apjl, 625, L123

\bibitem[{{Jenet} {et~al.}(2006){Jenet}, {Hobbs}, {van Straten}, {Manchester},
  {Bailes}, {Verbiest}, {Edwards}, {Hotan}, {Sarkissian}, \& {Ord}}]{jenet06a}
{Jenet}, F.~A., {Hobbs}, G.~B., {van Straten}, W., {Manchester}, R.~N.,
  {Bailes}, M., {Verbiest}, J.~P.~W., {Edwards}, R.~T., {Hotan}, A.~W.,
  {Sarkissian}, J.~M., \& {Ord}, S.~M. 2006, \apj, 653, 1571

\bibitem[{{Johnston} {et~al.}(1993){Johnston}, {Lorimer}, {Harrison}, {Bailes},
  {Lyne}, {Bell}, {Kaspi}, {Manchester}, {D'Amico}, \&
  {Nicastro}}]{johnston93a}
{Johnston}, S., {Lorimer}, D.~R., {Harrison}, P.~A., {Bailes}, M., {Lyne},
  A.~G., {Bell}, J.~F., {Kaspi}, V.~M., {Manchester}, R.~N., {D'Amico}, N., \&
  {Nicastro}, L. 1993, \nat, 361, 613

\bibitem[{{Kargaltsev} {et~al.}(2004){Kargaltsev}, {Pavlov}, \&
  {Romani}}]{kargaltsev04a}
{Kargaltsev}, O., {Pavlov}, G.~G., \& {Romani}, R.~W. 2004, \apj, 602, 327

\bibitem[{{Loinard} {et~al.}(2007){Loinard}, {Torres}, {Mioduszewski},
  {Rodr{\'{\i}}guez}, {Gonz{\'a}lez-L{\'o}pezlira}, {Lachaume}, {V{\'a}zquez},
  \& {Gonz{\'a}lez}}]{loinard07a}
{Loinard}, L., {Torres}, R.~M., {Mioduszewski}, A.~J., {Rodr{\'{\i}}guez},
  L.~F., {Gonz{\'a}lez-L{\'o}pezlira}, R.~A., {Lachaume}, R., {V{\'a}zquez},
  V., \& {Gonz{\'a}lez}, E. 2007, \apj, 671, 546

\bibitem[{{Ma} {et~al.}(1998){Ma}, {Arias}, {Eubanks}, {Fey}, {Gontier},
  {Jacobs}, {Sovers}, {Archinal}, \& {Charlot}}]{ma98a}
{Ma}, C., {Arias}, E.~F., {Eubanks}, T.~M., {Fey}, A.~L., {Gontier}, A.-M.,
  {Jacobs}, C.~S., {Sovers}, O.~J., {Archinal}, B.~A., \& {Charlot}, P. 1998,
  \aj, 116, 516

\bibitem[{{Pradel} {et~al.}(2006){Pradel}, {Charlot}, \&
  {Lestrade}}]{pradel06a}
{Pradel}, N., {Charlot}, P., \& {Lestrade}, J.-F. 2006, \aap, 452, 1099

\bibitem[{{Shklovskii}(1970)}]{shklovskii70a} 
{Shklovskii}, I.~S. 1970, Soviet Astronomy, 13, 562 

\bibitem[{{Titov}(2007)}]{titov07a}
{Titov}, O.~A. 2007, Astronomy Letters, 33, 481

\bibitem[{{van Straten} {et~al.}(2001){van Straten}, {Bailes}, {Britton},
  {Kulkarni}, {Anderson}, {Manchester}, \& {Sarkissian}}]{van-straten01a}
{van Straten}, W., {Bailes}, M., {Britton}, M., {Kulkarni}, S.~R., {Anderson},
  S.~B., {Manchester}, R.~N., \& {Sarkissian}, J. 2001, \nat, 412, 158

\bibitem[{{Verbiest} {et~al.}(2008){Verbiest}, {Bailes}, {van Straten},
  {Hobbs}, {Edwards}, {Manchester}, {Bhat}, {Sarkissian}, {Jacoby}, \&
  {Kulkarni}}]{verbiest08a}
{Verbiest}, J.~P.~W., {Bailes}, M., {van Straten}, W., {Hobbs}, G.~B.,
  {Edwards}, R.~T., {Manchester}, R.~N., {Bhat}, N.~D.~R., {Sarkissian}, J.~M.,
  {Jacoby}, B.~A., \& {Kulkarni}, S.~R. 2008, \apj, 679, 675

\bibitem[{{Williams} {et~al.}(2004){Williams}, {Turyshev}, \&
  {Boggs}}]{williams04a}
{Williams}, J.~G., {Turyshev}, S.~G., \& {Boggs}, D.~H. 2004, Physical Review
  Letters, 93, 261101

\bibitem[{{Zakamska} \& {Tremaine}(2005)}]{zakamska05a}
{Zakamska}, N.~L., \& {Tremaine}, S. 2005, \aj, 130, 1939

\bibitem[{{Zavlin} {et~al.}(2002){Zavlin}, {Pavlov}, {Sanwal}, {Manchester},
  {Tr{\"u}mper}, {Halpern}, \& {Becker}}]{zavlin02a}
{Zavlin}, V.~E., {Pavlov}, G.~G., {Sanwal}, D., {Manchester}, R.~N.,
  {Tr{\"u}mper}, J., {Halpern}, J.~P., \& {Becker}, W. 2002, \apj, 569, 894

\end{thebibliography}

\end{document}